\documentclass[12pt,a4paper,epsf]{article}
\usepackage{graphics}
\usepackage{amssymb,amsmath}
\usepackage[dvips]{lscape,graphicx}
\usepackage{cite}
\usepackage{longtable}
\usepackage{slashed}
\usepackage{bbold}

\newcommand{\ct}{\cite}
\newcommand{\lb}{\label}

\newcommand{\bc}{\begin{center}}
\newcommand{\ec}{\end{center}}
\newcommand{\bd}{\begin{displaymath}}
\newcommand{\ed}{\end{displaymath}}
\newcommand{\be}{\begin{equation}}
\newcommand{\ee}{\end{equation}}
\newcommand{\ba}{\begin{array}}
\newcommand{\ea}{\end{array}}
\newcommand{\bea}{\begin{eqnarray}}
\newcommand{\eea}{\end{eqnarray}}
\newcommand{\bt}{\begin{tabular}}
\newcommand{\et}{\end{tabular}}
\newcommand{\un}{\underline}

\newcommand{\bp}{\begin{picture}}
\newcommand{\ep}{\end{picture}}
\newcommand{\bfi}{\begin{figure}}
\newcommand{\efi}{\end{figure}}

\def\fun#1#2{\lower3.6pt\vbox{\baselineskip0pt\lineskip.9pt
\ialign{$\mathsurround=0pt#1\hfil##\hfil$\crcr#2\crcr\sim\crcr}}}

\begin{document}

\vspace{0.5cm}

\title{\LARGE \bf {Gravi-Weak Unification and the Black-Hole-Hedgehog's Solution with Magnetic
Field Contribution}}
\author{\large \bf
  B.G. Sidharth ${}^{1}$\footnote{iiamisbgs@yahoo.co.in, birlasc@gmail.com},\,\,
  C.R. Das ${}^{2}$\footnote{das@theor.jinr.ru},\,\,
  L.V. Laperashvili ${}^{3}$\footnote{laper@itep.ru},
  H.B. Nielsen${}^{4}$\footnote{hbech@nbi.dk}\\\\
{\large \it ${}^{1}$ International Institute of Applicable Mathematics}\\
{\large \it and Information Sciences,}\\
{\large \it B.M. Birla Science Centre}\\
{\large \it Adarsh Nagar, 500063 Hyderabad, India}\\\\
{\large \it ${}^{2}$ Bogoliubov Laboratory of Theoretical Physics}\\
{\large \it Joint Institute for Nuclear Research}\\
{\large \it International Intergovernmental Organization,}\\
{\large \it Joliot-Curie 6, 141980 Dubna, Moscow region, Russia}\\\\
{\large \it ${}^{3}$ The Institute of Theoretical and
Experimental Physics,}\\
{\large\it National Research Center ``Kurchatov Institute'',}\\
{\large\it Bolshaya Cheremushkinskaya, 25, 117218 Moscow, Russia}\\\\
{\large \it ${}^{4}$ Niels Bohr Institute,}\\
{\large \it Blegdamsvej, 17-21, DK 2100 Copenhagen, Denmark}}

\date{}
\maketitle

\thispagestyle{empty}

\vspace{2cm}



\clearpage \newpage

\thispagestyle{empty}

\newpage

\begin{abstract}

In the present paper, we investigated the gravitational
black-hole-hedgehog's solution with magnetic field contribution in
the framework of the f(R)--gravity described by the Gravi-Weak
unification model. Assuming the Multiple Point Principle (MPP), we
considered the existence of the two degenerate vacua of the
Universe: the first Electroweak (EW) vacuum with $v_1 \approx 246$
GeV (``true vacuum"), and the second Planck scale (``false vacuum")
with $v_2 \sim 10^{18}$ GeV. In these vacua, we investigated
different topological defects. The main aim of this paper is an
investigation of the black-hole-hedgehog configurations as defects
of the ``false vacuum". We have obtained the solution which
corresponds to a global monopole, that has been ``swallowed" by the
black-hole with core mass $M_{BH}\approx 3.65\times 10^{18}\,\,
{\rm{GeV}}$ and radius $\delta \approx 6\cdot 10^{-21}
{\rm{GeV}}^{-1}.$ We investigated the metric in the vicinity of
the black-hole-hedgehog and estimated its horizon radius:
$r_h\approx 1.14 \delta$. We have considered the phase transition
from the ``false vacuum" to the ``true vacuum" and confirmed the
stability of the EW--vacuum.

\end{abstract}

\vspace{2cm}

\section{Introduction}

In the previous papers, \ct{1,2,3} devoted to studying of
topological defects of the universal vacua we gave the
investigation of hedgehog's configurations \ct{1*,2*} neglecting
the contribution of magnetic fields.

The birth of our Universe is a Big Bang since it represents the
point of time when the Universe entered into a regime where the
laws of physics began to work. Big Bang is not an explosion in
space, but rather an expansion of space. After the initial
expansion, the early Universe underwent a series of phase
transitions. During these phase transitions, the breakdown of
local or global gauge symmetries produces the vacuum topological
defects (point, line and sheet defects). The cosmological model
developed in Refs.~\ct{1,2,3} assumes the existence of two
degenerate vacua of the Universe: The first (``true") Electroweak
(EW) vacuum with VEV $v_1 \approx 246$ GeV, and the second
(``false") Planck scale vacuum with VEV $v_2 \sim 10^{18}$ GeV. In
these papers, we investigated hedgehog's configurations as defects
of ``the false vacuum".

Gravitational black holes solutions with Yang-Mills fields were
investigated explicitly in Ref.~\ct{1x}. But in the present paper
we investigate a solution for a global monopole, which is a
black-hole-hedgehog at the Planck scale in the framework of the
f(R)-gravity predicted by the Gravi-Weak unification (GWU) model,
previously developed by authors in Refs.~\ct{4,5,6,7}. In contrast
to the theory \ct{1,2,3}, here we have taken into account the
contribution of the magnetic field of hedgehogs.

\section{Gravi-Weak unification, the action and field equations}

In Refs.~\ct{4,5,6,7}, using results of Refs.~\ct{8,9}, we have
constructed the Gravi-Weak unification (GWU), considering a
$Spin(4, 4)$-group of GWU spontaneously broken into the
$SL(2,C)^{(grav)}\times SU(2)^{(weak)}$ group of symmetry. In
agreement with experimental and astrophysical results, we assumed
that after the Big Bang, there came into being the unification
group $G_{TOE}$ of the {\bf Theory of the Everything (TOE)} which
was rapidly broken down to the direct product of series of gauge
groups (see Ref.~\ct{1}) ended by the Standard Model group
$G_{SM}$:
\be GSM = SU(3)_c \times SU(2)_L \times U(1)_Y . \lb{1} \ee
The action $S_{(GW)}$ of the Gravi-Weak unification obtained in
Refs.~\ct{4,5,6,7} is given by the following expression:
\bea S_{(GW)} &=& - \frac 1{g_{uni}}\int_{\mathfrak
M}d^4x\sqrt{-g}\left[\frac 1{16}\left(R|\Phi|^2 - \frac 32|\Phi|^4\right)\right.\nonumber\\
&&+\left. \frac 1{16}\left(aR_{\mu\nu}R^{\mu\nu} + bR^2\right) +
\frac 12{\cal D}_{\mu}\Phi^\dag{\cal D}^{\mu}\Phi + \frac 14
F_{\mu\nu}^iF^{i\,\mu\nu}\right],  \lb{15} \eea
where $g_{uni}$ is a parameter of the GWU, parameters $a,b$ (with
$a+b=1$) are ``bare" coupling constants of the higher derivative
gravity, $R$ is the Riemann curvature scalar, $R_{\mu\nu}$ is the
Ricci tensor, $|\Phi|^2 = \Phi^a\Phi^a$ is a squared triplet Higgs
field, where $\Phi^a$ (with $a=1,2,3$) is an isovector scalar
belonging to the adjoint representation of the $SU(2)$ gauge group
of symmetry. In Eq.(\ref{15}):
\be   {\cal D}_{\mu}\Phi^a = \partial_{\mu}\Phi^a +
g_2\epsilon^{abc}A_{\mu}^b\Phi^c       \lb{16} \ee
is a covariant derivative, and
\be   F_{\mu\nu}^a = \partial_{\mu}A_{\nu}^a -
\partial_{\nu}A_{\mu}^a + g_2\epsilon^{abc}A_{\mu}^bA_{\nu}^c  \lb{17} \ee
is a curvature of the gauge field $A_{\mu}^a$ of the $SU(2)$
Yang-Mills theory with a coupling constant $g_2$ as a ``bare"
coupling constant of the $SU(2)$ weak interaction.

The action (\ref{15}) is a special case of the $f(R)$ gravity
\ct{10,11,12}  when:
\be f(R) = R|\Phi|^2.    \lb{18} \ee
General case of the $f(R)$ gravity gives the action containing
matter fields and can be presented by the following expression:
\be S = \frac{1}{2\kappa}\int d^4x \sqrt{-g}\,f(R) + S_{grav} +
S_{gauge} + S_m, \lb{19} \ee
where the action $S_m$ is associated with matter fields (fermions
and Higgs fields).

From the action (\ref{15}), using the metric formalism, we obtain
the following field equations:
\be {\cal F}(R)R_{\mu\nu} - \frac 12 f(R)g_{\mu\nu} - \nabla_{\mu}
\nabla_{\nu}{\cal F}(R) + g_{\mu\nu}\Box {\cal F}(R) = \kappa
T_{\mu\nu}^m,                             \lb{20} \ee
where:
\be {\cal F}(R)\equiv \frac{df(r)}{dr}|_{r=R},    \lb{20a} \ee
$\kappa = 8\pi G_N$, $G_N$ is the gravitational constant, and
$T^m$ is the energy-momentum tensor derived from the matter action
$S_m$.

Varying the fields $\Phi$ and $A_{\mu}$, we obtain the next field
equations:
\be {\cal D}_{\mu}{\cal D}^{\mu}\Phi = (\frac{R}8 - \lambda
|\Phi|^2) \Phi,   \lb{21} \ee
where according to Ref~\ct{1} (see Appendix A, Eq.~(A13)), we
have: $$\lambda = \frac{3g_2^2}{8},$$ and
\be {\cal D}^{\mu}F_{\mu\nu} = - J_{\nu}, \lb{22} \ee
where $J_{\nu}$ is a current, produced by the Higgs field
$\Phi^a$:
\be  J_{\nu} = \frac 12[\Phi^\dagger {\cal D}_{\nu}\Phi - ({\cal
D}_{\nu}\Phi^\dagger)\Phi].  \lb{23} \ee

\section{De Sitter solutions at the early time of the Universe}

It is well known that at the early time, the Universe is described
by the de Sitter solutions (see for example Refs.~\ct{13,14}). Our
model is a special case of the more general SU(N) model \ct{8},
where authors assumed that the Universe is inherently de Sitter.
Then, the 4-spacetime is a hyperboloid in a 5-dimensional
Minkowski space under the constraint:
\be  x^2_0 + x^2_1 + x^2_2 + x^2_3 + x^2_4 = r^2_{dS}, \lb{24} \ee
where $r_{dS}$ is a radius of the curvature of the de Sitter
space, or simply ``the de Sitter radius".

Vacuum energy density of our Universe is the Dark Energy (DE). The
cosmological constant $\Lambda$ describes the DE substance, which
is dominant in the Universe at later times:
\be  \Omega_{DE} = \frac{\rho_{DE}}{\rho_{crit}} \simeq 0.75,
                                         \lb{25} \ee
where $\rho_{DE}$ is the dark energy density and the critical
density is:
\be  \rho_{crit} = \frac{3H_0^2}{8\pi G_N}\simeq 1.88\times
10^{-29}H_0^2.   \lb{26} \ee
Here $H_0$ is the Hubble constant:
\be H_0 \simeq 1.5 \times 10^{-42}\,\, {\rm {GeV}}.  \lb{27} \ee
Dark Energy (DE) is related with cosmological constant $\Lambda$
by the following way:
\be \rho_{DE} = \rho_{vac} = (M^{red}_{Pl})^2\Lambda,  \lb{28} \ee
where $M^{red}_{Pl}$ is the reduced Planck mass:
$M^{red}_{Pl}\simeq 2.43 \times 10^{18}$ GeV.

At present, cosmological measurements give (see \ct{PDG}):
\be  \rho_{DE} \simeq (2\times 10^{-3}\,\, {\rm{ eV}})^4,  \lb{29}
\ee
which means a tiny value of the cosmological constant:
\be \Lambda\simeq 10^{-84}\,\, {\rm GeV}^2. \lb{30} \ee
This tiny value of $\rho_{DE}$ was first predicted by B.G.
Sidharth in 1997 year \ct{1a,2a}. In the 1998 year S. Perlmutter,
B. Schmidt and A. Riess \ct{3a} were awarded the Nobel Prize for
the discovery of the accelerating expansion of the Universe.

Having an extremely small cosmological constant of our Universe,
Bennett, Froggatt and Nielsen \ct{BN,BFN,FN} assumed to consider
only zero, or almost zero, cosmological constants for all vacua
existing in Nature. They formulated a new law of Nature named
{\un{the Multiple Point Principle (MPP)}}, which means: {\it There
exist in Nature several degenerate vacua with very small energy
density, or cosmological constants.}

The model developed in this article considers the existence of the
two degenerate vacua of the Universe: The first (``true")
Electroweak (EW) vacuum, and the second (``false") Planck scale
vacuum.

From experimental results, cosmological constants -- minima of the
Higgs effective potentials $V_{eff} (\phi_H)$ -- are not exactly
equal to zero. Nevertheless, they are extremely small. By this
reason, Bennett, Froggatt and Nielsen \ct{BN,BFN,FN} assumed to
consider zero cosmological constants as a good approximation. Then
according to the MPP, we have a model of pure SM being fine-tuned
in such a way that these two vacua proposed have just zero energy
density (see also Ref.~\ct{LV}).

If the effective potential has two degenerate minima, then the
following requirements are satisfied:
\be V_{eff} (\phi^2_{min1}) =  V_{eff} (\phi^2_{min2}) = 0,
\lb{6a} \ee
and \be V'_{eff} (\phi^2_{min1}) =  V'_{eff} (\phi^2_{min2}) = 0,
\lb{7a} \ee
where
\be V'(\phi^2) = \frac{\partial V}{\partial \phi^2}. \lb{8a} \ee
Here we assume that:
\be V_{eff} (\phi^2_{min1}) = V_{EW}, \quad V_{eff}
(\phi^2_{min2}) = V_{high\,\, field}.     \lb{9a} \ee
Assuming the existence of the two degenerate vacua in the SM:

a) the first Electroweak vacuum at $v_1 \approx 246$ GeV, and

b) the second Planck scale vacuum at $v_2 \sim 10^{18}$ GeV,\\
Froggatt and Nielsen predicted in Ref.~\ct{FN} the top-quark and
Higgs boson masses:
\be M_t = 173 \pm 5\,\, {\rm { GeV}}; \quad M_H = 135 \pm 10\,\,
{\rm { GeV}}. \lb{10a} \ee
In the present paper we study the evolution of the Universe as two
bubbles: one having a ``false vacuum", and the other one having a
``true vacuum". The bubble, which we shall refer to as the {\bf
false vacuum}, to be a de Sitter space with a constant expansion
rate $H_F$. This bubble has a radius close to the de Sitter
horizon, which corresponds to the Universe radius:
\be R_{un}\simeq R_{de\,\, Sitter\,\, horizon}\simeq 10^{28}\,\,
{\rm{cm}}.  \lb{31} \ee
It is convenient to use the flat de Sitter coordinates to describe
the background of the inflating false vacuum:
\be ds^2 = dt^2 - e^{2H_F t} (dr^2 + r^2d\Omega^2),  \lb{32} \ee
where
\be   d\Omega^2 = d\theta^2 + \sin^2\theta d\phi^2.   \lb{33} \ee
The space-time inside the bubble, which we shall refer to as a
{\bf true vacuum}, has the geometry of an open
Friedmann-Lemaitre-Robertson-Walker (FLRW) universe (see for
example review \ct{Cop}):
 \be ds^2 = d\tau^2 - a(\tau)^2(d\xi^2 + \sinh^2\xi d\Omega^2),
\lb{34} \ee
where $a(\tau)$ is a scale factor with cosmic time $\tau$. In the
true vacuum we have a constant expansion rate $H_T$, which has the
meaning of the slow-roll inflation rate inside the bubble at the
early stage of its evolution.

Cosmological theory of bubbles was developed in a lot of papers by
A. Vilenkin and his collaborators (see for example,
Refs.~\ct{15,16,17,18}).

As it was shown in Ref.\ct{8}, the nontrivial vacuum solution to
the action (\ref{15}) is de Sitter spacetime with a non-vanishing
Higgs vacuum expectation value (VEV) of the triplet Higgs scalar
field $\Phi$: $v_2 = \langle\Phi\rangle = \Phi_0$. The standard
Higgs potential in Eq.(\ref{15}) has an extremum at $\Phi_0 = R/3$
(with $R
> 0$), corresponding to a de Sitter spacetime background solution:
\be   R = R_0 =\frac{12}{r_{dS}^2} = 3v_2^2,
                                   \lb{35} \ee
which implies vanishing curvature:
\be F_0 = \frac 12 R_0 - \frac 1{16}\Sigma_0\Phi_0^2   \lb{36} \ee
solving the field equations $DF = dF + [A, F] = 0$, and strictly
minimizing the action (\ref{15}).

Based on this picture, the origin of the cosmological constant
(and DE) is associated with the inherent spacetime geometry, and
not with vacuum energy of particles (we consider their
contributions later). We note that as a fundamental constant under
the de Sitter symmetry, $r_{dS}$ is not a subject to quantum
corrections. Local dynamics exist as fluctuations with respect to
this cosmological background. In general, the de Sitter space may
be inherently unstable. The quantum instability of the de Sitter
space was investigated by various authors. Abbott and Deser
\ct{19} have shown that de Sitter space is stable under a
restricted class of classical gravitational perturbations. So any
instability of the de Sitter space may likely have a quantum
origin. Ref.\ct{20} demonstrated through the expectation value of
the energy-momentum tensor for a system with a quantum field in a
de Sitter background space, that in general, it contains a term
that is proportional to the metric tensor and grows in time. As a
result, the curvature of the spacetime would decrease and the de
Sitter space tends to decay into the flat space (see
Ref.~\ct{21}). The decay time of this process  is of the order of
the de Sitter radius:
\be   \tau \sim r_{dS}\simeq 1.33\,\, H_0^{-1}.  \lb{37} \ee
Since the age of our universe is smaller than $\tau_{dS}$, we are
still observing the accelerating expansion of the Universe.

Of course, we also can consider the perturbation de Sitter
solutions but these perturbations are very small
\ct{13,14}.\\

\section{The solution for the gravitational black-holes-hedgehogs with magnetic
field contribution}

The field configurations describing a monopole-hedgehog \ct{1*,2*}
are:
\be \Phi^a = v w(r )\frac{x^a}{r},  \lb{1m}  \ee
\be A_{\mu}^a = a(r )\epsilon_{\mu ab}\frac{x^b}{r},  \lb{2m} \ee
where $x^ax^a = r^2$ with $(a = 1, 2, 3)$, $w(r)$ and $a(r)$ are
some structural functions. This solution is pointing radially.
Here $\Phi^a$ is parallel to $\hat{r}$ -- the unit vector in the
radial, and we have a ``hedgehog" solution of Refs.~\ct{1*,2*}. The
terminology ``hedgehog" was first suggested by Alexander Polyakov
in Ref.~\ct{2*}.

In the flat metric, the field equations (\ref{21}) for $\Phi^a$
give the following equation for $w(r)$:
\be  w'' + \frac 2{r}w' - \frac 2{r^2}w - \frac{w(w^2
-1)}{\delta^2} - 2 w a^2(r) = 0,     \lb{3m} \ee
where $\delta$ is a core radius of the hedgehog.

The function $w(r)$ grows with $r$ from $w(0)=0$ and exponentially
approaches to the unity: $\lim w(r)_{r \to \infty} = 1$ . Barriola
and Vilenkin \ct{15} took $w = 1$ outside the core when $r\geq
\delta$, which is an approximation to the exact solution. As a
result, the functions $w(r)$ and $a(r)$ are constrained by the
following conditions:
\be  w(0) = 0, \quad {\rm{and}} \quad w(r)\to 1 \quad
{\rm{when}}\quad r\to \infty,    \lb{5m} \ee
\be a(0) = 0, \quad {\rm{and}} \quad a(r)\sim - \frac {g}{r} \quad
{\rm{when}}\quad r \to \infty.    \lb{6m} \ee

\subsection{The metric in the vicinity of the global monopole}

The most general static metric in the vicinity of the global
monopole is a metric with a spherical symmetry:
\be ds^2 = B(r)dt^2 - A(r)dr^2 -r^2(d\theta^2 + sin^2\theta
d\varphi^2).     \lb{7m} \ee
For this metric the Ricci tensor has the following non-vanishing
components:
$$ R_{tt} = -\frac{B''}{2A} + \frac{B'}{4A}\left(\frac{A'}{A} +
\frac{B'}{B}\right) - \frac{1}{r}\frac{B'}{A},$$
$$  R_{rr} = \frac{B''}{2B} + \frac{B'}{4B}\left(\frac{A'}{A} +
\frac{B'}{B}\right) - \frac{1}{r}\frac{A'}{A},$$
$$  R_{\theta\theta} = - 1 + \frac{r}{2A}\left( - \frac{A'}{A} +
\frac{B'}{B}\right) + \frac{1}{A},$$ \be R_{\varphi\varphi} =
\sin^2\theta R_{\theta\theta}.   \lb{8m} \ee
Now we can calculate the global monopole energy-momentum tensor
components:
$$  T^t_t = v^2\frac{w'^2}{2A} + v^2\frac{w^2}{r^2} + \frac 14
\lambda v^4 (w^2 - 1)^2 -\frac{a'^2}{A} + \frac{a^2}{r^2}, $$
$$  T^r_r = - v^2\frac{w'^2}{2A} + v^2\frac{w^2}{r^2} + \frac 14
\lambda v^4 (w^2 - 1)^2 -\frac{a'^2}{A} + \frac{a^2}{r^2},
$$ \be T^\theta_\theta = T^\varphi_\varphi = v^2\frac{w'^2}{2A} +
\frac 14 \lambda v^4 (w^2 - 1)^2.       \lb{9m} \ee
Here $\kappa=1$.

\subsection{The hedgehog's structure functions}

As an example we can use the following expressions for monopole
structure functions $w(r)$ and $a(r)$, which satisfy the
conditions (\ref{5m}) and (\ref{6m}):
\be w(r) = 1 - \exp(-\frac{r^2}{\delta^2}),\quad \lb{10m} \ee
\be a(r) = - \frac{g}{r}(1 - \exp(-\frac{r^2}{\delta^2})),\quad
\lb{11m} \ee
Then we see that in the vicinity of $r\to 0$ we have:
\be w(r) = \frac{r^2}{\delta^2} + ...  \quad {\rm{for}} \quad r\to
0,
 \lb{12m} \ee
and
\be a(r) = - g\frac{r}{\delta^2} + ... \quad {\rm{for}} \quad r\to
0, \lb{13m} \ee
But in the vicinity of $r\to \infty$ we have:
\be w(r) \to 1 - ... \quad {\rm{for}} \quad r\to \infty,\lb{14m}
\ee
and
\be a(r) \to -\frac{g}{r} + ... \quad {\rm{for}} \quad r\to
\infty,\lb{15m} \ee
in accordance with the conditions (\ref{5m}) and (\ref{6m}).

Of course, we are able to calculate the components (\ref{9m}) of
the monopole energy-momentum tensor using the result (\ref{10m})
and (\ref{11m}). But for simple estimation we can be limited by an
approximation:
\be w(r) = 0 \quad {\rm{for}} \quad 0\leq r\leq \delta, \quad
{\rm{and}} \quad
 w(r) = 1\quad {\rm{for}}\quad \delta < r < \infty, \lb{16m} \ee
and
\be a(r) = 0, \quad g=0\quad {\rm{for}}\quad 0\leq r\leq \delta,
\quad {\rm{and}} \quad a(r) = -\frac{g}{r},\quad g=g_2\quad
{\rm{for}} \quad \delta < r < \infty, \lb{17m} \ee
where $\delta$ is a radius of the hedgehog.

\subsection{The solution with magnetic field contribution}

Considering the approximation (\ref{16m}) and (\ref{17m}), in
agreement with a solution used by Barriola and Vilenkin in
Ref.~\ct{15}, we can obtain a simple approximate solution for the
monopole-hedgehog taking $w=1$ out the core of the hedgehog
\ct{1}. In the case of Refs.~\ct{1,35,36,37,ShiLi,Car} scalar
curvature $R$ is constant, and Eq.~(\ref{20}) comes down to the
Einstein's equation:
\be  \frac{1}{A}\left(\frac{1}{r^2} -
\frac{1}{r}\frac{A'}{A}\right) - \frac{1}{r^2} = \frac{1}{\kappa
v^2}T^t_t,   \lb{18m} \ee
\be  \frac{1}{A}\left(\frac{1}{r^2} +
\frac{1}{r}\frac{B'}{B}\right) - \frac{1}{r^2} = \frac{1}{\kappa
v^2}T^r_r.   \lb{19m} \ee
Here $\kappa = 8\pi G_N$.

In approximation (\ref{16m}) and (\ref{17m}), the energy-momentum
tensor components are given by the following approximations:
\be T^t_t = T^r_r \approx  \frac{\lambda \kappa^2 v^4}{4}  \quad
{\rm{for}} \quad 0\leq r\leq \delta, \lb{20m} \ee
\be T^t_t = T^r_r \approx \frac{\kappa v^2}{r^2} +
\frac{g^2}{r^4}(- \frac{1}{A} + 1)   \quad {\rm{for}}\quad \delta
< r < \infty, \lb{21m} \ee
By substraction of Eqs.(\ref{18m}) and (\ref{19m}) we obtain:
\be    \frac{A'}{A} + \frac{B'}{B} = 0. \lb{24m} \ee
From Eq.~(\ref{18m}) we obtain a general relation for the function
$A(r)$:
\be   A^{-1}(r) = 1 - \frac{1}{r}\int_0^r T^t_t\, r^2dr. \lb{25m}
\ee
Using expressions (\ref{20m}) and (\ref{21m}), we obtain:
\be   A^{-1}(r) = 1 - \frac{\lambda \kappa^2 v^4}{4r}\int_0^\delta
r^2 dr  - \frac{1}{r}\int_\delta^r \big(\frac{\kappa v^2}{r^2} +
\frac{g^2}{r^4}(- \frac{1}{A} + 1)\big )r^2dr. \lb{26m} \ee
Taking into account the Schwarzschild type metric for a black-hole
given by Refs.~\ct{1,1x,A}, we can use the following expression
for $A^{-1}(r)$:
\be   A^{-1}(r) \approx C + \frac{C_1}{r} + \frac{C_2}{r^2} +
\frac{C_3}{r^3} + ... ,  \lb{27m} \ee
and from Eq.~(\ref{26m}) we calculate:
\be
 C = 1 - \kappa v^2, \quad C_1 = \kappa v^2 \delta,
\quad C_2 = g^2 \kappa v^2,\quad C_3 = - \frac{g^2}2\kappa v^2
\delta, ... \lb{28m}. \ee
Eq.~(\ref{24m}) gives:
\be  A(r) = B^{-1}(r),   \lb{29m} \ee
and finally we obtain the following result:
\be  A^{-1}(r) = B(r) \approx 1 - \kappa v^2 + \frac{\kappa v^2
\delta}{r} + \frac{g^2\kappa v^2}{r^2} - \frac{g^2\kappa
v^2\delta}{2r^3} + ... \lb{30m} \ee
According to Refs.~\ct{1,1x,A}, the Schwarzschild type metric for
a black-hole is given by the expression:
\be  A^{-1} \approx 1  - \kappa v^2 - \frac{2G_N\,M}{r} + ...,
\lb{31m} \ee
where $M$ is a black-hole's mass parameter, which in our theory is
given by the following expression:
\be M \approx - 4\pi v^2\delta. \lb{32m} \ee
The mass parameter (\ref{32m}) is negative. There is a repulsive
gravitational potential due to this negative mass parameter given
by the metric parameter $A(r)$. But this parameter $M$ is not a
mass of the hedgehog: the black-hole-hedgehog has a positive mass:
\be M_{BH}= - M = 4\pi v^2 \delta.\lb{33m} \ee

If we take the space integral of the hedgehog energy density as
given by (\ref{20m}, \ref{21m}), say - this would be total energy
of the hedgehog in the ignoring gravity approximation - the
integration over the radius $r$ will diverge for large $r$,
because of the term $\frac{\kappa v^2}{r^2}$ in (\ref{21m}). So
with the approximations done the a priori ``mass" = ``energy" of the
hedgehog is $+\infty$. The {\em positivity} is what one expects
for a disturbance in a background vacuum, which has minimum energy
density and the {\em divergence} comes from the kinetic term due
to the variation of the $\Phi^a$ field because of having different
directions in a component space essentially following the
direction in space from the center out. Indeed such a variation
leads to a gradient square term behaving $\propto \frac{1}{r^2}$.
When this is integrated over space - meaning an integral $\int
...4\pi r^2 dr$ one gets a term proportional to upper end $r$ and
thus divergence. It is a kind of infrared divergence in the sense
that it comes from large distance scales.

This is a priori looking like a hedgehog-``soliton" having an
infinite energy or mass. And it is needed to give some comments on
it and its understanding:

\begin{itemize}
\item In fact, this divergence causing contribution is not really
there if we calculate fully in our Gravi-Weak theory in as far as
the contribution is showing up as just a gauge artefact if we
include the gauge field associated with the transformation of
components of the $\Phi^a$ scalar field. As one sees from the
Lagrangian the kinetic term for the $\Phi^a$ field involves {\em
covariant derivatives}, and then one can arrange a lower energy
density by letting the Yang-Mills fields adjust to make the actual
values of these covariant derivatives in the far-out surroundings
of the hedgehog be zero. As a price for this vanishing of the
covariant derivatives one has to accept the magnetic fields as
they then come, but this gives crudely more differentiation and a
dependence with a more negative power of $r$ ensuring convergence
in the large $r$ limit.

So really this problem of divergence - if we consider it a problem
- is due to having ignored the Yang-Mills field.

\item As one may note from the performed calculation the divergent
causing term goes into the r-independent term $C$ in the expansion
(\ref{27m}) of the inverse of $A$, rather than into the term going
as $C_1/r$ as one would expect for a mass. One could look at this
phenomenon as the constant coefficient $C$ in the $A^{-1}$
expansion somehow ``renormalizing'' the mass so as to take up in
itself the divergent part.

\item It is only because of there being such
``renormalization-like'' trick going on in the calculation that it
can at all be possible to circumvent the a priori expectation that
``the mass $M$" must, of course, be positive because of the hedgehog
being an excitation on the background of a ground state.

But once one can interpret part of the a priori mass contribution
as something else by shuffling it into the constant $C$ term in
the (\ref{27m}) expansion, of course even the sign of the rest -
which then is interpreted indeed as a mass $M$ - gets out of
control and there is no contradiction by it being negative.

\end{itemize}

\section{Lattice--like structure of the false vacuum}

Now we can construct  the lattice-like topological contribution
with negative vacuum energy density.

Assuming that black-holes with mass parameter $M=-M_{BH}$ form a
hypercubic lattice with lattice parameter $l=\lambda_{Pl}$, we
have the negative energy density (and negative cosmological
constant $\Lambda_{lat}$) of such a lattice equal to:
\be    \rho_{lat} \simeq - M_{BH}M_{Pl}^3 = \Lambda_{lat}M_{Pl}^2.
\lb{34m} \ee
If this energy density of the hedgehogs lattice compensates the
Einstein's vacuum energy (see (A10) and (A15)), we have the
following equation:
\be   \frac{\lambda}{4}v^4 \approx M_{BH}M_{Pl}^3.
                                     \lb{35m} \ee
Using the estimation (A8), we obtain:
\be \frac 32 M_{Pl}^4 \approx M_{BH} M_{Pl}^3, \lb{36m} \ee
or
\be M_{BH} = \frac 32 M_{Pl}\approx 3.65\times 10^{18}\,\,
{\rm{GeV}}. \lb{37m} \ee
Therefore black-holes-hedgehogs have a huge mass of order of the
Planck mass.

Eq.(\ref{33m}) predicts a radius $\delta$ of the hedgehog's core:
\be \delta \approx \frac{M_{BH}}{4\pi v^2}\approx
\left(\frac{64\pi}{3} M_{Pl}\right)^{-1}\approx 6\cdot
10^{-21}\,\, {\rm{GeV}}^{-1}.              \lb{38m} \ee

\subsection{The hedgehog's horizon radius}

We have obtained a global monopole with a huge mass (\ref{37m}).
This is a black-hole solution, which corresponds to a global
monopole-hedgehog that has been ``swallowed" by a black-hole.
Indeed, we have obtained the metric result by M. Barriola et al.
\ct{15} like:
\be ds^2 = \left(1 - \kappa v^2 + \frac{2G_N\,M_{BH}}{r}
+...\right)dt^2 - \frac{dr^2}{(1 - \kappa v^2 +
\frac{2G_N\,M_{BH}}{r} + ...)} - r^2(d\theta^2 + \sin^2\theta
d\varphi^2). \lb{39m} \ee
A black hole has a horizon. A horizon radius $r_h$ can be found by
solving the equation:
\be A^{-1}(r_h) = 0. \lb{40m} \ee
Assuming that $g^2=0$ in the region $r\leq r_h$, we obtain:
 \be ( 1 - \kappa v^2)(1 + \frac{2G_N M_{BH}}{(1-\kappa v^2)r_h}) = 0. \lb{41m} \ee
Eq.~(\ref{41m}) gives a solution for hedgehog's horizon radius:
\be  r_h \approx  \frac{\kappa\,M_{BH}}{4\pi (\kappa v^2 - 1)}.
\lb{42m} \ee
According to Eq.~(A7), $\kappa v^2 = 8$, and the
black-hole-hedgehog's horizon radius is equal to:
\be  r_h \approx  \frac{\frac{\kappa}{4\pi}\times M_{BH}}{1 -
\kappa v^2} = \frac{\kappa v^2 \delta}{1 - \kappa v^2} \approx
\frac{8}{7}\delta \approx 1.14\delta. \lb{43m} \ee
We see that the horizon radius $r_h$ is larger than the hedgehog
radius $\delta$:
$$r_h > \delta,$$
and our concept that ``a black hole contains the hedgehog" is
justified.

\subsection{Lattice-like structure of the false vacuum and
non-commutativity}

We see that at the Planck scale the false vacuum of the Universe
is described by a non-differentiable space-time: by a foam of
black-holes, having lattice-like structure, in which sites are
black-holes with ``hedgehog" monopoles inside them. This manifold
is described by a non-commutative geometry (see Ref.~\ct{1}).

 In Refs.~\ct{1a,2a} B.G. Sidharth predicted:
\begin{enumerate}
\item[1.] That a cosmological constant is given by a tiny value:
\be  \Lambda \sim H_0^2,   \lb{15a} \ee
where $H_0$ is the Hubble rate in the early Universe:
\be  H_0 \simeq 1.5 \times 10^{-42}\,\, {\rm GeV}.  \lb{16a} \ee
\item[2.] That a Dark Energy density is very small:
\be  \rho_{DE} \simeq 10^{-12}\,\, {\rm eV}^4 = 10^{-48}\,\, {\rm
GeV}^4;  \lb{17a} \ee
\item[3.] That a very small DE-density provides an accelerating
expansion of our Universe after the Big Bang.
\end{enumerate}

Sidharth proceeded from the following points of view \ct{44}:
Modern Quantum Gravity \ct{Rov} (Loop Quantum Gravity, etc.,) deal
with a non-differentiable space-time manifold. In such an
approach, there exists a minimal space-time cut off
$\lambda_{min}$, which leads to the non-commutative geometry.

If the space-time is fuzzy, non-differentiable, then it has to be
described by a non-commutative geometry with the coordinates
obeying the following commutation relations:
\be  [dx^{\mu}, dx^{\nu}] \approx \beta^{\mu\nu}l^2 \neq 0.
\lb{20a} \ee
Eq.~(\ref{20a}) is true for any minimal cut off $l$.

Previously the following commutation relation was considered by
H.S. Snyder \ct{snyd}:
\be   [x, p] = \hslash \left( 1 + \left(\frac{l}{\hslash}
\right)^2p^2\right),\,\, etc.,  \lb{21a} \ee
which shows that effectively 4-momentum $p$ is replaced by
\be   p \to  p\left( 1 +
\left(\frac{l}{\hslash}\right)^2p^2\right)^{-1}.
                           \lb{22a} \ee
Then the energy-momentum formula becomes as:
\be E^2 \approx m^2 + p^2 - 2\left(\frac{l}{\hslash}\right)^2p^4.
\lb{24a} \ee
In such a theory the usual energy momentum dispersion relations
are modified. In the above equations, $l$ stands for a minimal
(fundamental) length, which could be the Planck length
$\lambda_{Pl}$, or for more generally -- Compton wavelength
$\lambda_c$.

Writing Eq.~(\ref{24a}) as
\be    E = E' + E'',   \lb{25a} \ee
where $E'$ is the usual (old) expression for energy, and $E''$ is
the new additional term in modification. $E''$ can be easily
verified as $E'' = - m_bc^2$ -- for boson fields, and $E'' =
+m_fc^2$ -- for fermion fields with masses $m_b,m_f$,
respectively. These formulas help to identify the DE density, what
was first realized by B.G. Sidharth in Ref.~\ct{2a}.

DE density is a density of the quantum vacuum energy of the
Universe. Quantum vacuum, described by Zero Point Fields (ZPF)
contributions is the lowest state of any Quantum Field Theory
(QFT), and due to the Heisenberg's principle has an infinite
value, which is renormalizable.

As it was pointed out in Refs.~\ct{Zel,42}, the quantum vacuum of
the Universe can be a source of the cosmic repulsion. However, a
difficulty in this approach has been that the value of the
cosmological constant turns out to be huge \ct{Zel}, far beyond
the value which is observed by astrophysical measurements. This
phenomenon has been called ``the cosmological constant problem"
\ct{Wein}.

A global monopole is a heavy object formed as a result of the
gauge-symmetry breaking during the phase transition of the
isoscalar triplet $\Phi^a$ system. The black-holes-hedgehogs are
similar to elementary particles, because of a major part of their
energy is concentrated in a small region near the monopole core.
Assuming that the Planck scale false vacuum is described by a non-
differentiable space-time having lattice-like structure, where
sites of the lattice are black-holes with ``hedgehog" monopoles
inside them, we describe this manifold by a non-commutative
geometry with a minimal length $l=\lambda_{Pl}$.

The result (\ref{34m}) is in agreement with the result of the
non-commutativity (\ref{25a}) with $E'' = - M_{BH}c^2$, because
black-holes-hedgehogs are point-like topological defects similar
to scalar particles giving the negative contribution $\rho_{lat} <
0$ to the vacuum energy density $\rho_{DE}$ of the Universe.

Using the non-commutative theory of the discrete space-time, B.G.
Sidharth predicted in Refs.~\ct{2a,42} a tiny value of the
cosmological constant: $\Lambda \simeq 10^{-84}$ GeV$^2$ as a
result of the compensation of ZPF contributions by non-commutative
contributions of the (boson and fermion) lattices.

\section{The phase transition from the ``false vacuum" to the ``true vacuum"}

In the present model, we investigated the evolution of the two
bubbles of the Universe, considering two phases of the universal
vacua:

\begin{enumerate}
\item one being a ``false vacuum" (Planck scale vacuum), and \item
the other is a ``true vacuum" (EW--vacuum).
\end{enumerate}

The cosmological model predicts that the Universe exists in the Planck
scale phase for extremely short time. For this reason, the Planck
scale phase was called ``the false vacuum". The presence of
hedgehogs as vacuum defects is responsible for the destabilization
of the false vacuum. The decay of the false vacuum is accompanied
by the decay of the black-holes-hedgehogs. These configurations
are unstable, and at some finite cosmic temperature which is
called the critical temperature $T_c$, a system exhibits a
spontaneous symmetry breakdown, and we observe a phase transition
from the bubble with the false vacuum to the bubble with the true
vacuum. After the phase transition, the Universe begins its
evolution toward the low energy Electroweak (EW) phase. Here the
Universe underwent the inflation, which led to the phase having
the VEV $v_1\approx 246$ GeV. This is a ``true" vacuum, in which we
live.

Ref.~\ct{35} also allowed a possibility to consider an arbitrary
domain wall between these two phases. During the inflation, domain
wall annihilates, producing gravitational waves and a lot of the
SM particles, having masses.

The Electroweak spontaneous breakdown of symmetry $SU(2)_L\times
U(1)_Y \to U(1)_{el.mag}$ leads to the creation of the topological
defects of the EW--vacuum. They are the Abrikosov-Nielsen-Olesen
closed magnetic vortices (``ANO strings") of the Abelian Higgs
model \ct{48,49}, and Sidharth's Compton phase objects \ct{50,51}.
Then the Electroweak vacuum and high-field ``false vacuum" both
present the non-differentiable manifold, described by the
non-commutative geometry,  giving almost zero cosmological
constants $\Lambda_1$ and $\Lambda_2$ (see\ct{1}).

At the early stage, the Universe was very hot, but then it began
to cool down. Black-holes-monopoles (as bubbles of the vapour in
the boiling water) began to disappear. The temperature dependent
part of the energy density died away. In that case, only the
vacuum energy will survive. Since this is a constant, the Universe
expands exponentially, and an exponentially expanding Universe
leads to the inflation (see reviews \ct{Lin,1Lin}). While the
Universe was expanding exponentially, so it was cooling
exponentially. This scenario was called {\bf supercooling in the
false vacuum.} When the temperature reached the critical value
$T_c$, the Higgs mechanism of the SM created a new condensate
$\phi_{min 1}$, and the vacuum became similar to a superconductor,
in which the topological defects are magnetic vortices. The energy
of black-holes is released as particles, which were created during
the radiation era of the Universe, and all these particles
(quarks, leptons, vector bosons) acquired their masses $m_i$
through the Yukawa coupling mechanism $Y_f \bar \psi_f\psi_f\phi$.
Therefore, they acquired the Compton wavelength,
$\lambda_i=\hbar/m_ic$. Then according to the Sidharth's theory of
the cosmological constant, in the EW-vacuum we again have
lattice-like structures formed by bosons and fermions, and the
lattice parameters ``$l_i$" are equal to the Compton wavelengths:
$l_i =  \lambda_i = \hbar/m_ic$.

\section{Stability of the EW--vacuum}

Here we emphasize that due to the energy conservation law, the
vacuum density before the phase transition (for $T > T_c$) is
equal to the vacuum density after the phase transition (for $T <
T_c$), therefore we have:
\be  \rho_{vac}({\rm at\,\, Planck\,\, scale}) = \rho_{vac}({\rm
at\,\, EW\,\, scale}).     \lb{12pt} \ee
The analogous link between the Planck scale phase and EW phase was
considered in the paper \ct{50}. It was shown that the vacuum
energy density (DE) is described by the different contributions to
the Planck and EW scale phases. This difference is a result of the
phase transition. However, the vacuum energy densities (DE) of
both vacua are equal, and we have a link between gravitation and
electromagnetism via the Dark Energy. According to the last
equation (\ref{12pt}), we see that if $\rho_{vac}$ (at the Planck
scale) is almost zero, then $\rho_{vac}$ (at EW scale) also is
almost zero, and we have a triumph of the Multiple Point
Principle: we have two degenerate vacua with almost zero vacuum
energy density. Almost zero cosmological constants are equal: $$
\Lambda_1 = \Lambda_2\approx 0. $$

Now we have obtained that the EW--vacuum, in which we live, is
stable. The Planck scale vacuum cannot be negative:
$V_{eff}(min_1) = V_{eff}(min_2)$.

\section{Conclusions}

1.  In the present paper, we investigated the topological
structure of the universal vacua. Different phase transitions,
which were resulted during the expansion of the early Universe
after the Planck era, produced the formation of the various kind
of topological defects in vacua of the Universe. The aim of this
investigation is the consideration of the hedgehog configurations
as defects in the false vacuum. We have obtained a solution for a
black-hole in the region which contains a global monopole in the
framework of the $f(R)$ gravity, where $f(R)$ is a function of the
Ricci scalar $R$. Here we have used the results of the Gravi-Weak
unification (GWU) model. The gravitational field, isovector scalar
$\Phi^a$ with $a = 1, 2, 3$, produced by a spherically symmetric
configuration in the scalar field theory, is pointing radially:
$\Phi^a$ is parallel to $\hat{r}$ -- the unit vector in the radial
direction. In this GWU approach, we obtained a ``hedgehog" solution
(in Alexander Polyakov's terminology). We also showed that this is
a black-hole solution, corresponding to a global monopole that has
been ``swallowed" by a black-hole.

2. In contrast to the previous theory \ct{1,2,3}, here we have
taken into account the contribution of the magnetic field of
hedgehogs.

3. We were based on the discovery that a cosmological constant of
our Universe is extremely small, almost zero, and assumed a new
law of Nature which was named as a Multiple Point Principle (MPP).
The MPP postulates: {\it There are two vacua in the SM with the
same energy density, or cosmological constant, and both
cosmological constants are zero, or approximately zero.} We
considered the existence of the following two degenerate vacua in
the SM: a) the first Electroweak vacuum at $v_1 = 246$ GeV, which
is a ``true" vacuum, and b) the second ``false" vacuum at the Planck
scale with VEV $v_2 \sim 10^{18}$ GeV.

4. The bubble, which we refer to as ``the false vacuum", is a de
Sitter space with its constant expansion rate $H_F$. The initial
radius of this bubble is close to the de Sitter horizon, which
corresponds to the Universe radius. The space-time inside the
bubble, which we refer to as ``the true vacuum", has the geometry
of an open FLRW universe.

5. By solving the gravitational field equations we estimated the
black-hole-hedgehog's mass, radius $\delta$ and horizon radius
$r_h$. They are: $M_{BH}\approx 3.65\times 10^{18}$ GeV,  $\delta
\approx 6\cdot 10^{-21}$ GeV$^{-1}$
 and $r_h \approx 1.14 \delta$.

6. We estimated all parameters of the Gravi-Weak unification
model, which gave the prediction of the Planck scale false vacuum
VEV equal to $v = 2\sqrt 2 M_{Pl}^{red}\approx 6.28\times 10^{18}$
GeV.

7. We have shown, that the Planck scale Universe vacuum is
described by a non-differentiable space-time: by a foam of
black-holes, or by lattice-like structure, where sites are
black-holes with the ``hedgehog" monopoles inside them. This
manifold is described by a non-commutative geometry, leading to a
tiny value of cosmological constant $\Lambda\approx 0$.

8. Taking into account that the phase transition from the ``false
vacuum" to the ``true vacuum" is a consequence of the electroweak
spontaneous breakdown of symmetry $SU(2)_L\times U(1)_Y \to
U(1)_{el.mag}$, we considered topological defects of EW-vacuum:
the Abrikosov-Nielsen-Olesen closed magnetic vortices (``ANO
strings") of the Abelian Higgs model and Sidharth's Compton phase
objects. We showed that the ``true vacuum" (EW-vacuum) again is
presented by the non-differentiable manifold with non-commutative
geometry leading to an almost zero cosmological constant.

9. We considered that due to the energy conservation law, the
vacuum energy density before the phase transition is equal to the
vacuum energy density after the phase transition: $
\rho_{vac}({\rm at\,\, Planck\,\, scale}) = \rho_{vac}({\rm at\,\,
EW\,\, scale}).$ This result confirms the Multiple Point
Principle: we have two degenerate vacua $v_1$ and $v_2$ with an
almost zero vacuum energy density (cosmological constants). By
this consideration we confirmed the vacuum stability of the
EW-vacuum, in which we live. The Planck scale vacuum cannot be
negative because of the exact equality $V_{eff}(min_1) =
V_{eff}(min_2)$.

\section*{Acknowledgments}
LVL greatly thanks to the B.M. Birla Science Centre (Hyderabad,
India) and personally Prof. B.G. Sidharth, for hospitality,
collaboration and financial support. HBN wishes to thank the Niels
Bohr Institute for the status of professor emeritus and
corresponding support. CRD is thankful to Prof. D.I. Kazakov for
support.

\section*{Appendix A. Parameters of the Gravi-Weak unification model}

At the first stage of the evolution (before the inflation), the
Universe had the de Sitter spacetime -- maximally symmetric
Lorentzian manifold with a constant and positive background scalar
curvature $R$. Then we have the following relations from the
action (\ref{15}):

1) The vacuum expectation value $v_2$ -- the VEV of ``the false
vacuum" -- is given by the de Sitter scalar curvature $R$:
$$    v_2^2 = \frac{R}{3}. \quad \quad (A1) $$

2) At the Planck scale the squared coupling constant of the weak
interaction is:
$$     g_2^2 = g_{uni}.\quad  \quad  (A2) $$
The replacement:
$$   \frac{\Phi^a}{g_2} \to \Phi^a \quad \quad (A3) $$
leads to the following GW-action:
$$  S_{(GW)} = - \int_{\mathfrak M}d^4x\sqrt{-g}\left(\frac
{R}{16}|\Phi|^2 - \frac{3g_2^2}{32}|\Phi|^4 + \frac 12{\cal
D}_{\mu}\Phi^\dag{\cal D}^{\mu}\Phi + \frac {1}{4g_2^2}
F_{\mu\nu}^iF^{i\,\mu\nu}\right)  + {\rm{grav.\,\, terms}}\big).
\quad \quad(A4) $$
Now considering the VEV of the false vacuum as $v=v_2$, we have:
$$  v^2 = \frac{R}{3g_2^2}. \quad \quad (A5)  $$
The Einstein-Hilbert action of general relativity with the
Einstein's cosmological constant $\Lambda_E$ is given by the
following expression:
$$  S_{EH} = - \frac 1{\kappa}\int d^4x \sqrt{-g}\left(\frac {R}{2}
- \Lambda_E\right). \quad (A6)    $$

3) The comparison of the Lagrangian $L_{EH}$ with the Lagrangian
given by Eq.~(A4) near the false vacuum $v$ leads to the following
relation between the Newton's gravitational constant $G_N$ and
reduced Planck mass:
$$    {(M_{Pl}^{red})}^2 = {(8\pi G_N)}^{-1} = \frac {1}{\kappa} =
\frac{v^2}{8}.  \quad (A7)   $$

4) Then we have:
$$  v = 2\sqrt 2 M_{Pl}^{red}\approx 6.28\times 10^{18}\,\,
{\rm{GeV}},    \quad (A8)        $$
and
$$ \Lambda_E = \frac{3g_2^2}{4} v^2. \quad (A9)    $$
Eq.~(A7) gives:
$$ \frac{1}{\kappa}\Lambda_E = \frac{3g_2^2}{32}v^4. \quad (A10) $$

The coupling constant $g_2$ is a bare coupling constant of the
weak interaction, which coincides with a value of the constant
$g_2$ at the Planck scale. Considering the renormalization group
equation (RGE) for the SU(2) running constant
$\alpha_2^{-1}(\mu)$, where $\alpha_2 = g_2^2 /4pi $ (see
Refs.~\ct{6x,7x}), we can carry out an extrapolation of this rate
to the Planck scale, what leads to the following estimations:
$$  \alpha_2(M_{Pl}) \sim \frac{1}{50},\quad g_{uni} = g_2^2 = 4\pi
\alpha_2(M_{Pl}) \approx 4\pi \times 0.02 \approx 0.25. \qquad
(A11)  $$

\subsection*{Appendix A. Global monopole}

A global monopole is described by the part $L_h$ of the Lagrangian
$L_{(GW)}$ given by the action (A4), which contains the
$SU(2)$-triplet Higgs field $\Phi^a$, VEV of the second vacuum
$v_2=v$ and cosmological constant $\Lambda=\Lambda_E$:
$$   L_h =  - \frac{R}{16}|\Phi|^2 + \frac{3g_2^2}{32}|\Phi|^4 - \frac
12\partial_{\mu}\Phi^a\partial^{\mu}\Phi^a + \Lambda_E $$ $$  = -
\frac 12
\partial_{\mu}\Phi^a\partial^{\mu}\Phi^a +
\frac{\lambda}{4}\left(|\Phi|^2 - v^2\right)^2 +
\frac{\Lambda_E}{\kappa} - \frac{\lambda}{4}v^4 = - \frac 12
\partial_{\mu}\Phi^a\partial^{\mu}\Phi^a +
\frac{\lambda}{4}\left(|\Phi|^2 - v^2\right)^2. \quad \quad (A12)
$$
Here  we have:
$$    \lambda = \frac {3g_2^2}{8}.  \quad \quad (A13)  $$
Substituting in Eq.~(A13) the value $g_2^2\approx 0.25$ given by
Eq.~(A11), we obtain:
$$    \lambda \approx \frac{3}{32}.   \quad \quad (A14)  $$
Eq.~(A9) gives:
$$ \frac{\Lambda_E}{\kappa} = \frac{3g_2^2}{32}v^4 =
\frac{\lambda}{4}v^4,     \quad \quad (A15) $$
and in Eq.~(A12) we have the compensation of the Einstein's
cosmological term. Then
$$    L_h =  - \frac 12
\partial_{\mu}\Phi^a\partial^{\mu}\Phi^a + V(\Phi),
                              \quad \quad (A16)     $$
where the Higgs potential is:
$$  V(\Phi) = \frac{\lambda}{4}\left(|\Phi|^2 - v^2\right)^2.
\quad \quad (A17) $$
This potential has a minimum at $\langle|\Phi|\rangle_{min}=v$, in
which it vanishes:
$$ V\left(|\Phi|^2_{min}\right) = V'\left(|\Phi|^2_{min}\right) =
0,\quad \quad (A18) $$
in agreement with the MPP conditions (\ref{6a}) and (\ref{7a}).

\end{document}